\documentclass[aps,prb,preprint,showpacs,preprintnumbers,amsmath,amssymb]
{revtex4}
\usepackage{amsmath}
\usepackage{amssymb}
\usepackage{amsfonts}
\usepackage{color,array}
\usepackage{dsfont}
\usepackage{slashed}
\usepackage{graphicx}
\usepackage{graphics}
\usepackage{epsfig}
\usepackage{bbm,bm}
\usepackage{psfrag}
\usepackage{ulem}
\usepackage{url}

\begin{document}

\title{Pseudo-$\epsilon$ Expansion and Renormalized Coupling Constants
at Criticality}

\author{A. I. Sokolov}
\email{ais2002@mail.ru}
\author{M. A. Nikitina}
\affiliation{Department of Quantum Mechanics,
Saint Petersburg State University,
Ulyanovskaya 1, Petergof,
Saint Petersburg, 198504
Russia}
\date{\today}

\begin{abstract}
Universal values of dimensional effective coupling constants $g_{2k}$ that 
determine nonlinear susceptibilities $\chi_{2k}$ and enter the scaling 
equation of state are calculated for $n$-vector field theory within the 
pseudo-$\epsilon$ expansion approach. Pseudo-$\epsilon$ expansions for 
$g_6$ and $g_8$ at criticality are derived for arbitrary $n$. Analogous 
series for ratios $R_6 = g_6/g_4^2$ and $R_8 = g_8/g_4^3$ figuring in 
the equation of state are also found and the pseudo-$\epsilon$ expansion 
for Wilson fixed point location $g_{4}^*$ descending from the six-loop RG 
expansion for $\beta$-function is reported. Numerical results are presented 
for $0 \le n \le 64$ with main attention paid to physically important 
cases $n = 0, 1, 2, 3$. Pseudo-$\epsilon$ expansions for quartic and sextic 
couplings have rapidly diminishing coefficients, so Pad\'e resummation 
turns out to be sufficient to yield high-precision numerical estimates. 
Moreover, direct summation of these series with optimal truncation gives 
the values of $g_4^*$ and $R_6^*$ almost as accurate as those provided by 
Pad\'e technique. Pseudo-$\epsilon$ expansion estimates for $g_8^*$ and 
$R_8^*$ are found to be much worse than that for the lower-order couplings 
independently on the resummation method employed.

Numerical effectiveness of the pseudo-$\epsilon$ expansion approach in two
dimensions is also studied. Pseudo-$\epsilon$ expansion for $g_4^*$ 
originating from the five-loop RG series for $\beta$-function of 2D 
$\lambda\phi^4$ field theory is used to get numerical estimates for $n$ 
ranging from 0 to 64. The approach discussed gives accurate enough values 
of $g_{4}^*$ down to $n = 2$ and leads to fair estimates for Ising and 
polymer ($n = 0$) models.

\end{abstract}


\maketitle

\section{Introduction}

The behavior of physical systems in the vicinity of Curie point
is characterized by universal parameters such as critical exponents,
critical amplitude ratios, etc. Among universal quantities important role
is played by renormalized effective coupling constants $g_{2k}$ which
enter small magnetization expansion of free energy and fix critical
asymptotes of nonlinear susceptibilities:
\begin{equation}
F(z,m) - F(0,m) = {\frac{m^3 }{g_4}} \Biggl({\frac{z^2 }{2}} + z^4 + {\frac{%
g_6 }{{g_4^2}}} z^6 + {\frac{g_8 }{{g_4^3}}} z^8 + ... \Biggr),
\label{eq:1}
\end{equation}
\begin{eqnarray}
\chi_4 = {\frac{\partial^3M }{{\partial H^3}}} \Bigg\arrowvert_{H = 0} = -
24 {\frac{\chi^2}{m^3}} g_4,
\qquad \quad
\chi_6 = {\frac{\partial^5M}{{\partial H^5}}} \Bigg\arrowvert_{H = 0} =
720 {\frac{\chi^3}{m^6}}(8 g_4^2 - g_6), \\
\nonumber
\chi_8 = {\frac{\partial^7M}{{\partial H^7}}} \Bigg\arrowvert_{H = 0} =
- 40320 {\frac{\chi^4}{m^9}}(96 g_4^3 - 24 g_4 g_6 + g_8), \qquad \quad \quad
\label{eq:2}
\end{eqnarray}
where $z$ is dimensionless magnetization, $z = M \sqrt{g_4/m^{1 + \eta}}$,
renormalized mass $m \sim (T - T_c)^\nu$ being the inverse correlation length,
$\chi$ is a linear susceptibility, $\chi_4$, $\chi_6$ and $\chi_8 $ -- nonlinear
susceptibilities of fourth, sixth and eighth orders.

Nonlinear susceptibilities and critical equation of state, describing the
influence of ordering field upon the behavior of a system near $T_c$,
attract permanent attention of theorists for decades. Dimensionless
effective coupling constants $g_{2k}$ and free energy (effective action)
for the basic models of phase transitions were found by a number of
analytical and numerical methods
\cite{B77,BNM78,LGZ80,BB,B1,B2,TW,B3,Reisz95,
AS95,CPRV96,S96,BC96,ZLF,SOU,GZ97,Morr,BC97,GZJ98,BC98,PV98,PV,SO,S98,SOUK99,
PV2000,ZJ01,CHPV2001,CHPRV2001,CHPRV2002,CPRV2002,ZJ,PV02,TPV2003,BT2005,
S2005,PV07,BP2011}. Calculation of the universal critical values of $g_4$
and $g_6$ for three-dimensional Ising model showed that the
field-theoretical renormalization group (RG) approach in fixed dimensions
yields highly accurate numerical estimates for these quantities. For
example, the resummation of four- and five-loop RG expansions by means of
the Borel-transformation-based procedures gave the values for $g_6^{*}$,
which differ from each other by less than $0.5\%$ \cite{SOU, GZ97} while
use of the resummed three-loop RG expansion enabled one to achieve an
apparent accuracy no worse than $1.6\%$ \cite{ SOU, S98}. In principle,
this is not surprising since the field-theoretical RG approach proved to
be highly efficient when used to estimate critical exponents, critical
amplitude ratios, marginal dimensionality of the order parameter, etc. for
various three-dimensional models \cite{BNM78, LGZ80, AS95, GZJ98, PS2000,
CPV2000, ZJ01, ZJ, PV02}. Moreover, the field-theoretical RG technique
turns out to be powerful enough even in two dimensions: properly resummed
four-loop \cite {BNM78, LGZ80} and five-loop \cite {OS2000} RG expansions
lead to fair numerical estimates for the critical exponents and
renormalized coupling constant $g_6^{*}$ \cite{SO} of 2D Ising model and
give reasonable results for other exactly solvable 2D models
\cite{MSS89,COPS04,NS13}.

Since RG expansions are known to be divergent, in order to obtain proper
numerical estimates one needs in resummation methods. Most of those
nowadays being employed are based on Borel transformation which avoids
factorial growth of higher-order coefficients and paves the way to
converging iteration schemes. This transformation widely used in the
theory of critical phenomena has resulted in a great number of high
precision numerical estimates. On the other hand, there exists alternative
technique turning divergent series into more "friendly" ones in the sense
that the expansions this technique yields have smaller lower-order
coefficients and much slower growing higher-order ones than those of
original series. We mean the method of pseudo-$\epsilon$ expansion put
forward by B. Nickel many years ago (see Ref. 19 in the paper of Le
Guillou and Zinn-Justin \cite{LGZ80}).

Pseudo-$\epsilon$ expansion approach was shown to be rather effective when
used to estimate numerical values of critical exponents and some other
universal quantities characterizing critical behavior of three-dimensional
systems \cite{LGZ80, GZJ98, FHY2000, HID04, CP05}. In two dimensions,
where original RG series are shorter and more strongly divergent,
pseudo-$\epsilon$ expansion technique is also able to give good or
satisfactory results \cite{LGZ80, COPS04, S2005, NS13}. To obtain
numerical estimates from pseudo-$\epsilon$ expansions one has to apply a
resummation technique since corresponding series have growing higher-order
coefficients, i. e. remain divergent. However, in contrast to RG expansions
in fixed and $4-\epsilon$ dimensions, pseudo-$\epsilon$ expansions do not
require advanced resummation procedures based on Borel transformation.
As a rule, use of simple Pad\'e approximants \cite{FHY2000,COPS04,S2005,NS13}
or even direct summation with an optimal cut off \cite{NS13} turn out to be
sufficient to obtain fair numerical estimates.

In this paper, we study renormalized effective coupling constants and
corresponding universal ratios of three-dimensional $O(n)$-symmetric
systems within the frame of pseudo-$\epsilon$ expansion technique. The
pseudo-$\epsilon$ expansions ($\tau$-series) for renormalized coupling
constants $g_6$ and $g_8$ will be calculated on the base of four-loop and
three-loop RG expansions obtained earlier \cite{S98,SOUK99} for 3D
$n$-vector field theory of $\lambda\varphi^4$ type. Along with the
higher-order couplings, Wilson fixed point coordinate $g_4^*$ and
universal critical values of ratios $R_6 = g_6/g_4^2$, $R_8 = g_8/g_4^3$
will be found as series in $\tau$ up to $\tau^6$, $\tau^4$ and $\tau^3$
terms respectively. The pseudo-$\epsilon$ expansions obtained will be
analyzed for $n = 0$, $n = 1$, $n = 2$, and $n = 3$, i. e. for the systems
most interesting from the physical point of view, as well as for other
values of $n$ ranging up to 64. In this context, Ising, XY and Heisenberg
models, apart from their physical importance, may be considered as
testbeds for clarification of numerical accuracy of the pseudo-$\epsilon$
expansion machinery since today we have a big amount of alternative
estimates for renormalized effective couplings obtained within the
$\epsilon$-expansion technique, on the base of perturbative RG expansions
in physical dimensions and extracted from lattice calculations and
computer simulations. Renormalized quartic coupling for the
two-dimensional $n$-vector model will be also studied for various $n$ and
comparison of the results given by pseudo-$\epsilon$ expansion approach
with their field-theoretical and lattice counterparts will be made.
Numerical estimates will be deduced from the pseudo-$\epsilon$ expansions
by means of Pad\'e and, when necessary, Pad\'e-Borel resummation methods
as well as by direct summation. The latter approach will be applied under
the assumption that proper numerical results may be obtained by means of
truncating divergent pseudo-$\epsilon$ expansions at smallest terms, i. e.
applying the procedure true for asymptotic series.

The paper is organized as follows. In the next section the
pseudo-$\epsilon$ expansions for $g_4^*$, $g_6^*$, $g_8^*$, $R_6^*$, and
$R_8^*$ are derived from 3D RG series for general $n$. Section III
contains numerical estimates of Wilson fixed point location for 3D
O($n$)-symmetric systems with $0 \le n \le 64$ resulting from
pseudo-$\epsilon$ expansion for $g_4^*$. Sections IV and V deal with
renormalized sextic and octic couplings respectively presenting and
discussing numerical estimates for $R_6^*$ and $R_8^*$ at various $n$.
In Section VI the quartic coupling constant of two-dimensional $n$-vector
model is studied within the pseudo-$\epsilon$ expansion approach. The
last section is a summary of the results obtained.

\section{Pseudo-$\epsilon$ expansions for quartic, sextic and octic
coupling constants}

As is well known, the critical behavior of D-dimensional systems with
O($n$)-symmetric vector order parameters may be described by Euclidean
field theory with the Hamiltonian:
\begin{equation}
\label{eq:3}
H = \int d^{D}x \Biggl[{1 \over 2}( m_0^2 \varphi_{\alpha}^2
 + (\nabla \varphi_{\alpha})^2)
+ {\lambda \over 24} (\varphi_{\alpha}^2)^2 \Biggr] ,
\end{equation}
where $\varphi_{\alpha}$ is a real $n$-vector field, bare mass squared
$m_0^2$ being proportional to $T - T_c^{(0)}$, $T_c^{(0)}$ -- mean field
transition temperature. The $\beta$-function for the model (3) in three
dimensions have been calculated within the massive theory \cite{Guelph,BNM78}
with the propagator, quartic vertex and $\varphi^2$ insertion normalized in
a standart way:
\begin{eqnarray}
\label{eq:4}
G_R^{-1} (0, m, g_4) = m^2 , \qquad \quad {{\partial G_R^{-1}
(p, m, g_4)} \over {\partial p^2}}
\bigg\arrowvert_{p^2 = 0} = 1 , \\
\nonumber
\Gamma_R (0, 0, 0, m, g) = m^2 g_4, \qquad \quad
\Gamma_R^{1,2} (0, 0, m, g_4) = 1.
\end{eqnarray}
Starting from the six-loop 3D RG expansion for $\beta$-function \cite{AS95}, we
replace the linear term in this expansion with $\tau g_4$, calculate the Wilson
fixed point coordinate as series in $\tau$, and arrive to the following expression:
\begin{eqnarray}
\label{g4-tau} g_4^* &=& \frac{2\pi}{n+8}\biggl[\tau + {\tau^2 \over (n + 8)^2}
\biggl(6.074074074~n + 28.14814815 \biggr)
\nonumber \\
&+& {\tau^3 \over (n + 8)^4} \biggl(- 1.34894276~n^3 + 8.056832799~n^2
+ 44.73231547~n - 12.48684745 \biggr)
\nonumber \\
&+& {\tau^4 \over (n + 8)^6} \biggl( - 0.15564589~n^5 - 7.638021730~n^4
+ 100.0250844~n^3 + 679.8756744~n^2
\nonumber \\
&+& 1604.099837~n + 3992.366079 \biggr)
\nonumber \\
&-& {\tau^5 \over (n + 8)^8} \biggl(0.05123618~n^7 + 4.68103281~n^6
+ 80.8238429~n^5 - 176.369063~n^4
\nonumber \\
&+& 11347.4861~n^3 + 153560.921~n^2 + 646965.181~n + 963077.072 \biggr)\biggr]
\nonumber \\
&+& {\tau^6 \over (n + 8)^{10}} \biggl(-0.0234242~n^9 - 2.5301565~n^8
- 71.923926~n^7 + 1183.9160~n^6 + 59058.036~n^5
\nonumber \\
&+& 631059.29~n^4 + 3909462.7~n^3 + 17512239~n^2 + 50941121~n + 66886678 \biggr)\biggr].
\end{eqnarray}

Substituting this expansion into four-loop RG series for sextic coupling
constant $g_6$ \cite{S98, SOUK99}

\begin{eqnarray}
\label{g6}
g_6 &=&{\frac 9\pi }g_4^3\Biggl[ {\frac{n+26}{{27}}}-{\frac{17~n+226}{{81\pi
}}}g_4+(0.000999164~n^2+0.14768927~n+1.24127452)g_4^2  \nonumber \\
&&-~(-0.00000949~n^3+0.00783129~n^2+0.34565683~n+2.14825455)g_4^3\Biggr]
\end{eqnarray}

and into three-loop series for $g_8$ \cite{SOUK99}

\begin{eqnarray}
\label{g8}
g_8 = -{\frac{81 }{{2 \pi}}} g_4^4 \Biggl[ {\frac{n + 80 }{{81}}} - {\frac{%
81~n^2 + 7114~n + 134960 }{{13122 \pi}}} g_4 \qquad \qquad \qquad \qquad
\nonumber \\
+ (0.00943497~n^2 + 0.60941312~n + 7.15615323) g_4^2 \Biggr]. \qquad
\end{eqnarray}

we obtain:

\begin{eqnarray}
\label{g6-tau} g_6^*&=&{8(n + 26)\pi^2 \tau^3 \over 3(n + 8)^3}+ {\tau^4
\over (n + 8)^5}(181.308289~n^2 + 8340.18127~n + 26061.6043)
\nonumber \\
&+& {\tau^5 \over (n + 8)^7} \biggl(- 78.4778860~n^4 - 1875.40831~n^3 +
37813.2081~n^2 + 191806.512~n
\nonumber \\
&+& 257751.564 \biggr) - {\tau^6 \over (n + 8)^9} \biggl(10.616530~n^6 +
1095.9774~n^5 + 25502.145~n^4
\nonumber \\
&-& 179690.51~n^3 - 616717.23~n^2 + 2241880.8~n + 7427442.9 \biggr).
\end{eqnarray}

\begin{eqnarray}
\label{g8-tau} g_8^*&=& -{8(n + 80)\pi^3\tau^4 \over (n + 8)^4} + {\tau^5
\over (n + 8)^6} \biggl(248.050213~n^3 + 17743.2461~n^2
\nonumber \\
&+& 77514.1600~n + 1072066.90 \biggr) \ \ + {\tau^6 \over (n + 8)^8}
\biggl(1387.95229~n^4 + 197852.837~n^3
\nonumber \\
&+& 1715306.54~n^2 + 15922970.4~n + 8711448.94 \biggr). \ \
\end{eqnarray}

Since small magnetization expansion of free energy contains ratios of
renormalized coupling constants
\begin{eqnarray}
\label{R6-8}
R_6 = \frac{g_6}{{g_4}^2}, \qquad \quad  R_8 = \frac{g_8 }{{g_4}^3} \ \
\end{eqnarray}
rather than coupling constants themselves, it is reasonable to calculate
pseudo-$\epsilon$ expansions for these ratios as well. They are as follows

\begin{eqnarray}
\label{R6-tau} R_6^*&=& \frac{2(n + 26)\tau}{3(n + 8)} - \frac{\tau^2}{(n
+ 8)^3}(3.506172836~n^2 + 36.83950607~n + 315.6543210)
\nonumber \\
&+& \frac{\tau^3}{(n + 8)^5} \biggl(-0.18927773~n^4 + 6.51351435~n^3 +
396.321683~n^2 + 2777.67913~n
\nonumber \\
&+& 10998.4537 \biggr) - \frac{\tau^4}{(n + 8)^7} \biggl(0.06139199~n^6 +
8.4167873~n^5 + 227.14320~n^4
\nonumber \\
&+& 3434.7520~n^3 + 49684.392~n^2 + 283809.46~n + 691313.24 \biggr). \ \
\end{eqnarray}

\begin{eqnarray}
\label{R8-tau} R_8^*&=& -\frac{(n + 80)\tau}{(n + 8)} + \frac{\tau^2}{(n +
8)^3}(n^3 + 89.75308641~n^2 + 1854.716049~n + 11077.53086)
\nonumber \\
&-& \frac{\tau^3}{(n + 8)^5} \biggl(16.6736016~n^4 + 1111.20557~n^3 +
22512.7084~n^2
\nonumber \\
&+& 199142.427~n + 713156.705 \biggr). \ \
\end{eqnarray}

\section{Wilson fixed point location from the pseudo-$\epsilon$ expansion}

Let us find numerical estimates for the fixed point value of quartic
coupling constant at various $n$ resulting from the pseudo-$\epsilon$
expansion (5). Address first the cases $n=0$, $n=1$, $n=2$ and $n=3$ that
are known to correspond to physically realizable systems.
Pseudo-$\epsilon$ expansions for the critical value of $g_4$ we'll deal
with are as follows:

\begin{eqnarray}
g_4^* &=&\frac{\pi}{4}\biggl(\tau + 0.4398148148\tau^{2} - 0.003048547
\tau^{3} + 0.015229668\tau^{4}
\nonumber\\
 &-& 0.05740387\tau^{5} + 0.0622931\tau^{6}\biggr), \qquad n = 0.
\end{eqnarray}

\begin{eqnarray}
g_4^* &=&\frac{2\pi}{9}\biggl(\tau + 0.4224965707\tau^{2} +
0.005937107 \tau^{3} + 0.011983594 \tau^{4}
\nonumber\\
&-& 0.04123101\tau^{5} + 0.0401346\tau^{6}\biggr), \qquad n = 1.
\end{eqnarray}

\begin{eqnarray}
g_4^* &=& \frac{\pi}{5} \biggl(\tau + 0.4029629630 \tau^2 +
0.009841357 \tau^3 + 0.010593080 \tau^4
\nonumber\\
&-& 0.02962102\tau^5+0.0282146\tau^6\biggr), \qquad n = 2.
\end{eqnarray}

\begin{eqnarray}
g_4^* &=& \frac{2\pi}{11}\biggl(\tau + 0.3832262014\tau^2 + 0.010777962\tau^3
+ 0.009577837\tau^4
\nonumber\\
&-& 0.02146532\tau^5 + 0.0211675\tau^6\biggr), \qquad n = 3.
\end{eqnarray}

The expansions for universal value of $g_4$ were, in fact, analyzed
earlier employing Borel transformation based resummation procedures
\cite{LGZ80,GZJ98}, although series (5), (13), (14), (15), (16)
themselves, to our knowledge, have never been published. Here we resum
these expansions and their counterparts for other $n$ by means of Pad\'e
approximants [L/M], i. e. using rather simple approach. This technique
is quite suitable in our situation since, as seen from (13)-(16), the
pseudo-$\epsilon$ expansions have small higher-order coefficients. Pad\'e
approximant technique is widely known today (see, e. g. Ref. \cite{BGM}),
so we write down Pad\'e tables for all four cases without going into detail.
Two points, however, have to be mentioned. First, to make comparison of
our estimates with others more convenient, we present numerical results
for rescaled constant $g = g_4(n + 8)/{2\pi}$; the series staying in
brackets in (5), (13), (14), (15), (16) are precisely the pseudo-$\epsilon$
expansions for this constant. Second, Pad\'e approximants are constructed
for $g^*/\tau$, with factor $\tau$ having physical value $\tau = 1$ ignored.
Tables I, II, III and IV present Pad\'e triangles discussed.

It looks natural to adopt as a final estimate for $g^*$ the average over
two highest-order near diagonal Pad\'e approximants [3/2] and [2/3]. We
do so for all the cases of interest apart from $n = 0$ when one of working
approximants -- [3/2] -- has abnormally large higher-order coefficients
(18.1, 41.2, 46.5) preventing obtaining high-precision estimate;
corresponding number is marked in Table I with $^+$. In this case we accept
as a most reliable the value given by another Pad\'e approximant -- [2/3].
So, our pseudo-$\epsilon$ expansion estimates for $g^*$ are:
\begin{equation}
g^* = 1.423~~(n = 0),\quad g^* = 1.423~~(n = 1),\quad g^* = 1.410~~(n =
2),\quad g^* = 1.393~~(n = 3),
\end{equation}
Numbers (17) differ from their canonical six-loop RG counterparts only
in third or even in fourth ($n = 3$) decimal place. This looks rather
optimistic encouraging to work further with Pad\'e resummed pseudo-$\epsilon$
expansions. Moreover, the accuracy of numerical results given by these
series rapidly improves when dimensionality of the order parameter $n$ grows
up. To demonstrate this we present Pad\'e triangle for $n = 6$ (Table V).
As seen from Table V, for this (not so big) value of $n$ the numbers given
by approximants [4/1], [3/2], and [2/3] practically coincide with each other
and with 6-loop RG estimate $g^* = 1.3385$ \cite{S98}.

Note that, as seen from Tables I--V, the numbers given by lower-order diagonal
and near diagonal Pad\'e approximants [2/2], [2/1], [1/2] are also close to
asymptotic values of $g$. It means that the pseudo-$\epsilon$ expansion
approach generates not only numerically efficient but rapidly converging
iteration procedure.

The overall situation is illustrated by Table VI accumulating pseudo-$\epsilon$
expansion estimates of $g^*$ for $0 \le n \le 64$. Along with Pad\'e estimates
(second column) the numbers obtained by direct summation of pseudo-$\epsilon$
expansions are presented here (third column). Direct summation is performed
under the assumption that one can get best numerical estimates truncating
divergent pseudo-$\epsilon$ expansions by smallest terms, i. e. adopting the
procedure valid for asymptotic series. Pad\'e estimates presented are the
averages over those given by near symmetric approximants [2/3] and [3/2], apart
from the case $n = 0$ (see above) when the value given by another approximant
is accepted as a final estimate. Numerical values of $g^*$ resulting from
analysis of 6-loop RG series in 3 dimensions$\cite{GZJ98,S98,BT2005}$ (fourth
and fifth columns), obtained within the $\epsilon$ expansion approach
$\cite{PV2000,BT2005}$ (sixth column) and extracted from lattice calculations
$\cite{BC98,PV98}$ (LC) are also collected in Table VI for comparison.

Table VI clearly demonstrates that the values of $g^*$ obtained from
Pad\'e approximants and given by direct summation are very close to each
other and to alternative high-precision estimates. Even for $n = 1$ the
difference between numbers produced by pseudo-$\epsilon$ expansion and by
other advanced techniques, both field-theoretical and lattice, is of order
of 0.01. This may be considered as a strong argument in favor of high
numerical effectiveness of the pseudo-$\epsilon$ expansion approach.
Moreover, direct summation of series for $g^*$ generates an iteration
procedure which, being quite primitive, rapidly converges to asymptotic
values that are very close to most accurate estimates known today. In this
sense, the pseudo-$\epsilon$ expansion approach itself may be referred to
as some special resummation technique. To confirm or to disprove this
statement, the structure of pseudo-$\epsilon$ expansions for other
universal quantities and corresponding numerical estimates are to be
analyzed.

\section{Sextic coupling and universal ratio $R_6$ for various $n$}

Let us estimate further universal critical values of $g_6$ and $R_6$
within the pseudo-$\epsilon$ expansion approach. For Ising, XY and
Heisenberg models corresponding $\tau$-series read:

~~~~~n = 1:
\begin{eqnarray}
g_6^* = \frac{8\pi^2}{81}\tau^3 + 0.585667731 \tau^4 + 0.101488719 \tau^5-
0.0229712 \tau^6.
\end{eqnarray}
\begin{eqnarray}
R_6^* = 2\tau - 0.488340192\tau^2 + 0.240118863\tau^3 - 0.2150291 \tau^4.
\end{eqnarray}

~~~~~n = 2:
\begin{eqnarray}
g_6^* = \frac {28\pi^2}{375} \tau^3 + 0.434672000 \tau^4 + 0.077635850
\tau^5 - 0.0084506 \tau^6.
\end{eqnarray}
\begin{eqnarray}
R_6^* = \frac{28}{15} \tau - 0.403358025 \tau^2 + 0.181881784 \tau^3 -
0.1489055 \tau^4.
\end{eqnarray}

~~~~~n = 3:
\begin{eqnarray}
g_6^* = \frac {232\pi^2}{3993} \tau^3 + 0.327311986 \tau^4 + 0.057293963
\tau^5 - 0.0025831 \tau^6.
\end{eqnarray}
\begin{eqnarray}
R_6^* = \frac{58}{33} \tau - 0.343898118 \tau^2 + 0.143177750 \tau^3 -
0.1079237 \tau^4.
\end{eqnarray}
Expansions for $g_6^*$ are seen to have fast diminishing coefficients with
irregular signs. On the contrary, coefficients of $\tau$-series for $R_6^*$
decrease more slowly but these series are alternating. We'll concentrate on
the numerical values of $R_6^*$ which enters the scaling equation of state and
has been estimated for various $n$ within several field-theoretical and lattice
methods$\cite{GZ97,PV,SOUK99,PV2000,BT2005,BP2011}$. Since higher-order coefficients
of series (18) -- (23) are rather small we do not need in Borel transformation
killing factorial growth of coefficients and can process our series by means of
Pad\'e approximants or even perform their direct summation. To clear up to what
extent numerical results are sensitive to the summation procedure we find the
values of $R_6^*$ in four different ways. Namely, we estimate $R_6^*$

i) by means of Pad\'e summation of series for $R_6^*$,

ii) via Pad\'e summation of series for $g_6^*$ and use of the first relation (10),

iii) by direct summation of $R_6^*$ pseudo-$\epsilon$ expansion with optimal
truncation, and

iv) by optimally truncated direct summation (OTDS) of $\tau$-series for
$g_6^*$ and subsequent use of (10) with $g_4^*$ also found by OTDS.

As was expected, Pad\'e resummation turns out to be effective in our problem.
One can see this from Pad\'e triangles for $g_6^*$ and $R_6^*$ at $n = 1$
presented in Tables VII and VIII. We choose here the Ising limit as an
illustration not only because of its physical significance. More important
point is that under $n = 1$ $\tau$-expansions for $g_6^*$ and $R_6^*$ have
larger higher-order coefficients than those for $n > 1$ making Ising model
rather "unfriendly" for pseudo-$\epsilon$ expansion analysis.

Numerical results obtained for $0 \le n \le 64$ are presented in Table IX.
The second column contains universal values of $R_6$ given by Pad\'e
summation of corresponding $\tau$-series. In the third column the
estimates found via Pad\'e summation of $\tau$-series for $g_6^*$ are
collected. The numbers obtained by optimally truncated direct summation of
the series for $R_6^*$ and $g_6^*$ form fourth and fifth columns. Pad\'e
estimates reported in Table IX are those averaged over two near diagonal
approximants [2/1] and [1/2] for $R_6^*/\tau$ and $g_6^*/\tau^3$. When one of
them is spoiled by a pole close to 1 or has abnormally large higher-order
coefficients the value given by another approximant is accepted as a final
estimate; these numbers are marked with asterisks. The values of $R_6^*$
resulting from 3D RG series$\cite{GZ97,SOUK99,PV2000,BT2005}$, obtained
within $\epsilon$-expansion$\cite{PV2000,BT2005}$ and
$1/n$-expansion$\cite{BW73,PV}$ approaches and extracted from lattice
calculations (LC) are also presented in the Table.

The values of $R_6^*$ staying in second and fifth columns of Table IX are
seen to be very close to each other and to RG estimates for any $n$. This
fact may be understood keeping in mind the structure of pseudo-$\epsilon$
expansions for $R_6^*$ and $g_6^*$. The series for $R_6^*$ have small enough
and monotonically decreasing coefficients with alternating signs what makes
their summation by means of Pad\'e approximants efficient$\cite{BGM}$. The
coefficients of the series for $g_6^*$, on the contrary, have irregular
signs but their modulo decrease extremely rapidly and the last
coefficients are tiny (see, e. g. (18), (20), (22)). This obviously favors
direct summation. On the other hand, $\tau$-series for $g_6^*$ because of
fast decreasing coefficients are also suitable for Pad\'e summation. That
is why the numbers in the third column of Table IX are rather close to
their counterparts from the second and fifth columns. In such a situation
optimally truncated direct summation of $\tau$-series for $R_6^*$ having no
advantages looks as a crude procedure, at least when compared with others
just discussed. Nevertheless, it provides quite satisfactory results for
$n \ge 10$ and leads to fair estimates for physical values of $n$.

So, we see that the pseudo-$\epsilon$ expansion approach combined with
Pad\'e resummation technique is a powerful instrument for analysis of
effective sextic interaction at criticality. Moreover, even direct
summation, if properly performed, is able to provide high-precision
numerical estimates for the universal ratio $R_6^*$ at any $n$.

\section{Octic coupling: structure of $\tau$-series and numerical estimates}

In the case of renormalized octic coupling we have shorter
pseudo-$\epsilon$ expansions with much less favorable structure. This is
clearly seen from the series for $n = 1, 2, 3$ written below:

~~~~~n = 1:
\begin{eqnarray}
g_8^* = -\frac{8\pi^3}{81} \tau^4 + 2.19699337 \tau^5 + 0.616747712 \tau^6
\end{eqnarray}
\begin{eqnarray}
R_8^* = -9\tau + 17.8641975 \tau^2 - 15.8502213 \tau^3.
\end{eqnarray}

~~~~~n = 2:
\begin{eqnarray}
g_8^* = -\frac{41\pi^3}{625} \tau^4 + 1.300052605 \tau^5 + 0.490236460
\tau^6.
\end{eqnarray}
\begin{eqnarray}
R_8^* = -\frac{41}{5} \tau + 15.1539753 \tau^2 - 12.1064882 \tau^3.
\end{eqnarray}

~~~~~n = 3:
\begin{eqnarray}
g_8^* = -\frac{664\pi^3}{14641} \tau^4 + 0.830338865 \tau^5 + 0.360948746
\tau^6.
\end{eqnarray}
\begin{eqnarray}
R_8^* = -\frac{83}{11} \tau + 13.1303207 \tau^2 - 9.59044946 \tau^3.
\end{eqnarray}

The series for $R_8^*$ being alternating have big elder coefficients. That is
why to estimate this ratio we apply, along with Pad\'e resummation,
Pad\'e-Borel procedure. Higher-order coefficients of the expansions for $g_8^*$
are much smaller. These series are processed within Pad\'e technique on the
base of approximant [1/1], the only nontrivial and diagonal one existing for
$g_8^*/\tau^3$. Then the value of $R_8^*$ is estimated using the second relation
(10).

Numerical results thus obtained are presented in Table X, along with the estimates
of $R_8^*$ deduced from RG series in three dimensions$\cite{GZ97,SOUK99,BT2005}$,
found within the $\epsilon$-expansion$\cite{PV2000,BT2005}$ and
$1/n$-expansion$\cite{PV}$ approaches and extracted from lattice calculations.
As is seen, in the case of octic coupling numerical estimates turn out to be
much worse than those obtained for $g_4^*$ and $R_6^*$. Indeed, the numbers
given by pseudo-$\epsilon$ expansions resummed in three different ways are
strongly scattered, to say nothing about their marked deviation from estimates
yielded by alternative methods. This is true not only for $n = 0, 1, 2, 3$, but
even for $n$ as large as 64: the difference between various pseudo-$\epsilon$
expansion estimates exceeds here 20\%.

Of course, pronounced shortness and strong divergence of $\tau$-series for
$g_8^*$ and $R_8^*$ may be thought of as main sources of such a failure. There
exists, however, an extra moment making the situation quite unfavorable. The
point is that the series (9), (12) have unusual feature. Namely, when
$n \to \infty$ the first and second terms in these expansions compensate each
another diminishing their mutual contribution and increasing the role of
higher-order terms; analogous peculiarity was observed earlier for original RG
expansion of $g_8$$\cite{SOUK99}$. Since each of $\tau$-series (9), (12) is
short and possesses only one such key higher-order term numerical effectiveness
of pseudo-$\epsilon$ expansion turns out to be poor in this case. We believe
that calculation of the next terms in $\tau$-series for renormalized octic
coupling would considerably improve the situation, as it occurs in other
unfavorable cases$\cite{OS2000}$. To get longer $\tau$-series one needs,
however, longer RG expansion for $g_8$. Today such an expansion is known only
for the Ising model$\cite{GZ97}$.

\section{Renormalized quartic coupling constant in two dimensions}

Here we'll apply the pseudo-$\epsilon$ expansion technique to estimate the
critical values of quartic coupling constant for two-dimensional systems.
Along with physically interesting cases $n = 1$ and $n = 0$ studied
earlier$\cite{S2005,NS13}$ the models with $n \ge 2$ will be considered.
Although these models are known not to undergo phase transitions into
ordered state, Wilson fixed point location has been calculated for them
within field-theoretical 2D RG approach in five-loop
approximation$\cite{OS2000}$ and using $\epsilon$-expansions constrained
at $D = 0$ and $D = 1$$\cite{PV98,PV2000}$. Comparison of numerical
results obtained within these techniques with those given by
pseudo-$\epsilon$ expansions is believed to shed light on computational
power of the latter approach.

Pseudo-$\epsilon$ expansion for fixed point value of $g$ in 2$D$ for arbitrary
$n$ is as follows$\cite{NS13}$:

\begin{eqnarray}
\label{g-tau}
g* &=& \tau + {\tau^2 \over (n + 8)^2} \biggl( 10.33501055~n + 47.67505273 \biggr)
\nonumber \\
&+& {\tau^3 \over (n + 8)^4} \biggl(- 5.00027593~n^3 + 24.4708201~n^2
+ 253.297221~n + 350.808487 \biggr)
\nonumber \\
&+& {\tau^4 \over (n + 8)^6} \biggl( 0.088842906~n^5 - 77.270445~n^4
+ 45.052398~n^3 + 3408.2839~n^2
\nonumber \\
&+& 14721.151~n + 27649.346 \biggr) - {\tau^5 \over (n + 8)^8}
\biggl(- 0.00407946~n^7 - 0.305739~n^6
\nonumber \\
&+& 1464.58~n^5 + 11521.4~n^4 + 98803.3~n^3 + 794945~n^2 + 3146620~n
+ 4734120 \biggr).
\end{eqnarray}

We estimate $g*$ for various $n$ lying between 0 and 64 within Pad\'e and
Pad\'e-Borel resummation techniques and by optimally truncated direct
summation. In course of Pad\'e resummation the approximant [3/2] is used
apart from the cases when it has poles close to 1. The choice of
approximant [3/2] is quite natural since it is equivalent to diagonal 
approximant [2/2] for $g^*/\tau$, i. e. with insignificant factor $\tau$
neglected. Pad\'e-Borel resummation is based on highest-order approximants
having no positive axis ("dangerous") poles that prevent evaluation
of Borel integral. Direct summation is performed as before with truncation
on the term with smallest coefficient.

The results obtained are collected in Table XI. The fixed point values of $g$
resulting from 5-loop RG series$\cite{OS2000}$, obtained within constrained
$\epsilon$-expansion approach$\cite{PV2000}$ and extracted from
$(1/n)$-expansion$\cite{CPRV96}$ and lattice calculations$\cite{BC96,CHPV2001}$
(LC) are also presented there to compare with our data. As is seen from
Table XI pseudo-$\epsilon$ expansion results in quite good numerical estimates
for any $n$ provided Pad\'e or Pad\'e-Borel resummation is made. In fact,
use of Pad\'e-Borel resummation changes numerical estimates only slightly
leaving simple Pad\'e procedure effecient in two dimensions. Moreover, even
direct summation remains satisfactory at the quantitative level down to
$n = 4$ leading as well to reasonable numbers in physical cases $n = 1$ and
$n = 0$. So, estimating renormalized quartic coupling constant in two
dimensions on the base of pseudo-$\epsilon$ expansion one can use simplest
ways to process the series - Pad\'e approximants and direct summation.

These results lead us, as above, to the conclusion that the pseudo-$\epsilon$
expansion itself may be considered as a resummation method. The first argument
in favor of such a point of view is obvious: this approach turns strongly
divergent field-theoretical RG expansions into power series with smaller
lower-order coefficients and much slower increasing higher-order ones. The
second argument is specific for low-dimensional systems: the physical value
of the pseudo-$\epsilon$ expansion parameter $\tau$ is equal to 1, while the
Wilson fixed point coordinate $g^*$ playing analogous role within
field-theoretical RG approach is almost two times bigger for physical values of
$n$ ($g^*\approx 1.8$). This difference is essential, especially keeping
in mind importance of higher-order terms$\cite{OS2000}$.

\section{Conclusion}

To summarize, we have calculated pseudo-$\epsilon$ expansions for
universal values of renormalized coupling constants $g_4$, $g_6$, $g_8$ and
of ratios $R_6$, $R_8$ for 3D Euclidean $n$-vector $\lambda\phi^4$ field
theory. Numerical estimates for Wilson fixed point location $g_{4}^*$ and
for $R_6^*$ and $R_8^*$ have been found under $0 \le n \le 64$ using Pad\'e
and Pad\'e-Borel resummation techniques as well as by direct summation with
optimal truncation. For $g_{4}^*$ and $R_6^*$ pseudo-$\epsilon$ expansion
machinery was shown to lead to high-precision numerical estimates without
addressing Borel transformation. Moreover, in both cases properly performed
direct summation turned out to be sufficient to result in accurate enough
numbers at any $n$. This implies that the pseudo-$\epsilon$ expansion
approach itself may be thought of as some specific resummation technique.
For the octic coupling, however, this technique was shown to be much less
efficient: numerical estimates found by Pad\'e and Pad\'e-Borel summation
of $\tau$-series for $R_8^*$ and obtained via evaluation of $g_8^*$ are
strongly scattered and considerably deviate from their lattice and
field-theoretical counterparts. This failure, however, does not indicate
poor numerical effectiveness of the pseudo-$\epsilon$ expansion approach;
it is caused mainly by shortness of corresponding $\tau$-series and the
unfavorable feature of their structure.

Pseudo-$\epsilon$ expansion for renormalized quartic coupling constant of
2D $n$-vector field theory has been also analyzed. Universal values of $g_4$
for $0 \le n \le 64$ have been estimated using Pad\'e and Pad\'e-Borel
resummation techniques as well as by direct summation with optimal cut
off. Comparison of the results obtained with each other and with their
counterparts known from alternative field-theoretical and lattice
calculations has shown that pseudo-$\epsilon$ expansion technique provides
numerical estimates as accurate as those given by other advanced approaches.

\section*{Acknowledgment}

We dedicate our work to the memory of Kenneth Wilson whose talk at the
Soviet-American symposium in Leningrad in 1971 inspired one of us (A. I. S.)
to enter the Realm of Renormalization Group.

\newpage

\begin{table}[t]
\caption{Pad\'e table for pseudo-$\epsilon$ expansion of
quartic coupling constant $g^* = g_4^*(n+8)/{2\pi}$ at $n = 0$
(self-avoiding walks). Pad\'e approximants [L/M] are derived
for $g^*/\tau$, i. e. with factor $\tau$ omitted. Approximant
[3/2] has abnormally large higher-order coefficients (18.1, 41.2
and 46.5) preventing obtaining high-precision estimate;
corresponding number is marked with $^+$. The value of $g^*$
resulting from resummed original six-loop RG series and referred
to as most reliable RG estimate is equal to $1.413 \pm 0.006$
\cite{GZJ98}.}
\label{tab1}
\renewcommand{\tabcolsep}{0.4cm}
\begin{tabular}{{c}|*{6}{c}}
$M \setminus L$ & 0 & 1 & 2 & 3 & 4 & 5 \\ \hline
0 & 1      & 1.4398 & 1.4368 & 1.4520 & 1.3946 & 1.4569 \\
1 & 1.7851 & 1.4368 & 1.4393 & 1.4400 & 1.4245 & \\
2 & 1.3216 & 1.4512 & 1.4400 & 1.4394$^+$ & & \\
3 & 1.5298 & 1.4147 & 1.4226 & & & \\
4 & 1.3094 & 1.4240 & & & & \\
5 & 1.6014 & & & & & \\
\end{tabular}
\end{table}

\begin{table}[t]
\caption{Pad\'e triangle for pseudo-$\epsilon$ expansion of coupling
constant $g^* = g_4^*(n+8)/{2\pi}$ at $n = 1$ (Ising model). Pad\'e
approximants [L/M] are derived for $g^*/\tau$, i. e. with factor $\tau$
omitted. The value of $g^*$ given by resummed six-loop RG series is equal
to $1.411 \pm 0.004$\cite{GZJ98}.}
\label{tab2}
\renewcommand{\tabcolsep}{0.4cm}
\begin{tabular}{{c}|*{6}{c}}
$M \setminus L$ & 0 & 1 & 2 & 3 & 4 & 5 \\ \hline
0 & 1 & 1.4225 & 1.4284 & 1.4404 & 1.3992 & 1.4393 \\
1 & 1.7316 & 1.4285 & 1.4167 & 1.4311 & 1.4195 & \\
2 & 1.3332 & 1.4403 & 1.4313 & 1.4273 & & \\
3 & 1.4977 & 1.4118 & 1.4179 & & & \\
4 & 1.3373 & 1.4190 & & & & \\
5 & 1.5272 & & & & & \\
\end{tabular}
\end{table}

\begin{table}[t]
\caption{The same as Table II but for $n = 2$ (XY model). Approximant
[2/1] has a pole close to 1; its location is shown as a subscript.
Six-loop RG estimate for $g^*$ is $1.403 \pm 0.003$\cite{GZJ98}.}
\label{tab3}
\renewcommand{\tabcolsep}{0.4cm}
\begin{tabular}{{c}|*{6}{c}}
$M \setminus L$ & 0 & 1 & 2 & 3 & 4 & 5 \\ \hline
0 & 1 & 1.4030 & 1.4128 & 1.4234 & 1.3938 & 1.4220 \\
1 & 1.6749 & 1.4131 & 1.2741$_{0.93}$ & 1.4156 & 1.4082 & \\
2 & 1.3341 & 1.4235 & 1.4159 & 1.4119 & & \\
3 & 1.4674 & 1.4019 & 1.4072 & & & \\
4 & 1.3490 & 1.4082 & & & & \\
5 & 1.4791 & & & & & \\
\end{tabular}
\end{table}

\begin{table}[t]
\caption{The same as Table II but for $n = 3$ (Heisenberg model).
Approximant [2/1] has a pole close to 1; its location is shown as a
subscript. Six-loop RG estimate for $g^*$ is $1.390 \pm
0.004$\cite{GZJ98}.} \label{tab4}
\renewcommand{\tabcolsep}{0.4cm}
\begin{tabular}{{c}|*{6}{c}}
$M \setminus L$ & 0 & 1 & 2 & 3 & 4 & 5 \\ \hline 0 & 1 & 1.3832 & 1.3940 &
1.4036 & 1.3821 & 1.4033 \\
1 & 1.6213 & 1.3943 & 1.4800$_{1.13}$ & 1.3970 & 1.3928 & \\
2 & 1.3283 & 1.4037 & 1.3973 & 1.3943 & & \\
3 & 1.4383 & 1.3874 & 1.3922 & & & \\
4 & 1.3495 & 1.3934 & & & & \\
5 & 1.4422 & & & & & \\
\end{tabular}
\end{table}

\begin{table}[t]
\caption{The same as Table II but for $n = 6$. Approximant [2/1] has a
pole very close to 1; its location is shown as a subscript. The value of
$g^*$ given by Pad\'e-Borel resummed six-loop RG series is 1.3385
\cite{S98}.}
\label{tab5}
\renewcommand{\tabcolsep}{0.4cm}
\begin{tabular}{{c}|*{6}{c}}
$M \setminus L$ & 0 & 1 & 2 & 3 & 4 & 5 \\ \hline
0 & 1 & 1.3296 & 1.3362 & 1.3426 & 1.3335 & 1.3444 \\
1 & 1.4915 & 1.3363 & 1.5821$_{1.03}$ & 1.3389 & 1.3385 & \\
2 & 1.2946 & 1.3426 & 1.3390 & 1.3385 & & \\
3 & 1.3614 & 1.3353 & 1.33845 & & & \\
4 & 1.3200 & 1.3398 & & & & \\
5 & 1.3593 & & & & & \\
\end{tabular}
\end{table}

\begin{table}[t]
\caption{Fixed point values of quartic coupling constant $g$ for various
$n$ found by Pad\'e summation of corresponding pseudo-$\epsilon$
expansions and by direct summation of these series with the optimal cut
off (OTDS). Pad\'e estimates are averages over those given by near
diagonal approximants [2/3] and [3/2]. Since for $n = 0$ approximant
[3/2] has abnormally large higher-order coefficients (see caption to Table I)
the value given by another approximant is accepted as a final estimate; it
is marked by asterisk. The universal values of $g$ resulting from 6-loop
RG series in 3 dimensions$\cite{GZJ98,S98,BT2005}$, obtained within the
$\epsilon$-expansion approach$\cite{PV2000,BT2005}$ and extracted from lattice
calculations$\cite{BC98,PV98}$ (LC) are presented for comparison.}
\label{tab6}
\begin{tabular}{*{8}{c}}\hline
~~$n$~~~~ & ~~~Pad\'e~~~ & ~~~OTDS~~~ & ~~~3D RG$\cite{S98}$~~~ & ~~~3D RG~~~ &
~~~$\epsilon$-exp.$\cite{PV2000}$~~~ & ~~~LC$\cite{BC98}$~~~ & ~~~LC$\cite{PV98}$~~~ \\
\hline
0~~  & ~1.423$^*$~ & ~1.437~ & ~~ & ~1.413(6)$\cite{GZJ98}$~  & ~1.396(20)~ & ~1.388(5)~
& ~1.393(20)~ \\
1~~  & ~1.423~ & ~1.428~ & ~1.419~  & ~1.411(4)$\cite{GZJ98}$~ & ~1.408(13)~ & ~1.408(7)~
& ~1.406(9) ~ \\
2~~  & ~1.410~ & ~1.413~ & ~1.4075~ & ~1.403(3)$\cite{GZJ98}$~ & ~1.425(24) & ~1.411(8)~
& ~1.415(11)~ \\
3~~  & ~1.393~ & ~1.404~ & ~1.392~  & ~1.390(4)$\cite{GZJ98}$~ & ~1.426(9) & ~1.409(10)~
& ~1.411(12)~ \\
4~~  & ~1.375~ & ~1.383~ & ~1.3745~ & ~1.377(5)$\cite{GZJ98}$~ & ~1.393(21) & ~1.392(10)~
& ~1.396(16)~ \\
5~~  & ~1.357~ & ~1.362~ & ~1.3565~ & ~1.3569$\cite{BT2005}$~  & ~1.345$\cite{BT2005}$~ & & \\
6~~  & ~1.3385~ & ~1.3426~ & ~1.3385~ & ~1.3397$\cite{BT2005}$~  & ~1.321$\cite{BT2005}$~
& ~1.355(10)~ & \\
8~~  & ~1.3043~ & ~1.3024~ & ~1.3045~ & & ~1.307(6) & ~1.320(15)~ & ~1.321(10)~ \\
10~~ & ~1.2743~ & ~1.2733~ & ~1.2745~ & &  & ~1.290(15)~  &  \\
16~~ & ~1.2075~ & ~1.2090~ & ~1.2077~ & & ~1.202(4) & & ~1.215(5)~ \\
24~~ & ~1.1540~ & ~1.1542~ & ~1.1542~ & & ~1.150(4) & & ~1.158(4)~ \\
32~~ & ~1.1215~ & ~1.1216~ & ~1.1218~ & ~1.1219$\cite{BT2005}$~ & ~1.119(3) & & ~1.122(3) \\
40~~ & ~1.1001~ & ~1.1003~ & ~1.1003~ & & & &  \\
48~~ & ~1.0850~ & ~1.0852~ &  &  & ~1.085(2) &    & ~1.084(2)~ \\
64~~ & ~1.0652~ & ~1.0655~ &  & ~1.0656$\cite{BT2005}$~ & ~1.0638$\cite{BT2005}$~ & & \\
\hline
\end{tabular}
\end{table}

\begin{table}[t]
\caption{Pad\'e triangle for pseudo-$\epsilon$ expansion of
sextic coupling constant $g_6^*$ for Ising model.
Pad\'e approximants [L/M] are derived for $g_6^*/\tau^3$,
i. e. with factor $\tau^3$ omitted.}
\label{tab7}
\renewcommand{\tabcolsep}{0.4cm}
\begin{tabular}{{c}|*{4}{c}}
$M \setminus L$ & 0 & 1 & 2 & 3 \\ \hline
0 & 0.9748 & 1.5604 & 1.6619 & 1.6390 \\
1 & 2.4420 & 1.6832 & 1.6432 & \\
2 & 1.4858 & 1.6188 &  & \\
3 & 1.6582 & & & \\
\end{tabular}
\end{table}

\begin{table}[t]
\caption{Pad\'e triangle for pseudo-$\epsilon$ expansion of
universal ratio $R_6^*$ for Ising model. Pad\'e approximants [L/M]
are derived for $R_6^*/\tau$, i. e. with factor $\tau$ omitted.}
\label{tab8}
\renewcommand{\tabcolsep}{0.4cm}
\begin{tabular}{{c}|*{4}{c}}
$M \setminus L$ & 0 & 1 & 2 & 3 \\ \hline
0 & 2 & 1.5117 & 1.7518 & 1.5367 \\
1 & 1.6075 & 1.6726 & 1.6383 & \\
2 & 1.6896 & 1.6465 &  & \\
3 & 1.6036 & & & \\
\end{tabular}
\end{table}

\begin{table}[t]
\caption{Universal values of $R_6$ for various $n$ found by Pad\'e
summation of corresponding $\tau$-series (second column), obtained via
Pad\'e summation of $\tau$-series for $g_6$ (third column), given by
optimally truncated direct summation of the series for $R_6^*$ (fourth
column), and obtained via directly summed up $\tau$-series for $g_6^*$ and
$g_4^*$ (fifth column). Pad\'e estimates are those averaged over
approximants [2/1] and [1/2] for $R_6^*/\tau$ and $g_6^*/\tau^3$. If one of
them suffers from some pathology (see text) the estimate given by another
approximant is accepted as a final one; these numbers are marked with
asterisks. The values of $R_6^*$ resulting from 3D RG
series$\cite{GZ97,SOUK99,PV2000,BT2005}$, obtained within
$\epsilon$-expansion$\cite{PV2000,BT2005}$ and
$1/n$-expansion$\cite{BW73,PV}$ approaches and extracted from lattice
calculations (LC) are presented for comparison.} \label{tab9}
\begin{tabular}{*{9}{c}}\hline
$n$~~~~ & Pad\'e & Pad\'e & ~OTDS~ & ~OTDS for ~ & ~3D RG$\cite{SOUK99}$~ &
~3D RG~ & ~$\epsilon$-exp.$\cite{PV2000}$~ & LC and  \\
 &   &  for $g_6$  &   & $g_6$ and $g_4$ &   &   &  & $(1/n)$-exp. $(n \ge 8)$ \\ \hline
0~~ & ~1.726~ & ~1.733~ & ~1.556~ & ~1.727~ & ~~ & ~1.69(7)$\cite{PV2000}$~
& ~1.718(18)~ & ~~ \\
1~~ & ~1.642~ & ~1.654~ & ~1.537~ & ~1.649~ & ~1.648~ & ~1.644(6)$\cite{GZ97}$~ &
~1.652(15)~ & ~1.649$\cite{BP2011}$ \\
2~~ & ~1.566~ & ~1.576~ & ~1.496~ & ~1.574~ & ~1.574~ & ~1.576(10)$\cite{PV2000}$~ &
~1.575(10) & ~1.560(12)$\cite{CHPRV2001}$~ \\
3~~ & ~1.497~ & ~1.505~ & ~1.449~ & ~1.486~ & ~1.504~ & ~1.507(26)$\cite{PV2000}$~ &
~1.494(8) & ~1.49(3)$\cite{CHPRV2002}$~ \\
4~~ & ~1.436~ & ~1.439~ & ~1.401~ & ~1.427~ & ~1.442~ & ~1.447(22)$\cite{PV2000}$~ &
~1.424(7) & ~1.5(5)$\cite{Reisz95}$ \\
5~~ & ~1.381~ & ~1.384~ & ~1.355~ & ~1.375~ & ~1.387~ & ~1.38(2)$\cite{BT2005}$~ &
~1.36(1)$\cite{BT2005}$~ &  \\
6~~ & ~1.3325~ & ~1.335~ & ~1.312~ & ~1.327~ & ~1.338~ & ~1.33(2)$\cite{BT2005}$~ &
~1.31(2)$\cite{BT2005}$~ &  \\
8~~ & ~1.2505~ & ~1.2511~ & ~1.237~ & ~1.2554~ & ~1.254~ & ~~       & ~1.230(12) & ~1.688~ \\
10~~ & ~1.1849~ & ~1.1851~ & ~1.1752~ & ~1.1873~ & ~1.187~  &       &            & ~1.484~ \\
16~~ & ~1.0508~ & ~1.0506$^*$~ & ~1.0454~ & ~1.0567~ & ~1.050~ & ~~ & ~1.040(15) & ~1.177~ \\
24~~ & ~0.9506~ & ~0.9508$^*$~ & ~0.9464~ & ~0.9504~ & ~0.948~ & ~~ &            & ~1.007~ \\
32~~ & ~0.8899$^*$~ & ~0.8883~ & ~0.8877~ & ~0.8912~ & ~0.889~ & ~0.8885(6)$\cite{BT2005}$~ &
~0.889(8) & ~0.922 \\
40~~ & ~0.8510$^*$~ & ~0.8504~ & ~0.8525~ & ~0.8519~ & ~0.848~ &    &            & ~0.871~ \\
48~~ & ~0.8236$^*$~ & ~0.8234~ & ~0.8245~ & ~0.8243~ & ~~      & ~~ & ~0.823(4)  & ~0.837~ \\
64~~ & ~0.7877$^*$~ & ~0.7876~ & ~0.7879~ & ~0.7879~ &     &
~0.7855(3)$\cite{BT2005}$~ & ~0.7877(3)$\cite{BT2005}$~ & ~0.794~ \\
\hline
\end{tabular}
\end{table}

\begin{table}[t]
\caption{The values of $R_8^*$ for various $n$ found by Pad\'e
summation of corresponding pseudo-$\epsilon$ expansions (second column),
by Pad\'e-Borel summation of these series (third column) and obtained via
Pad\'e summation of the series for $g_8^*$ (fourth column). The values of
$R_8^*$ resulting from RG series in 3 dimensions$\cite{GZ97,SOUK99,BT2005}$, obtained
within the $\epsilon$-expansion$\cite{PV2000,BT2005}$ and $1/n$-expansion$\cite{PV}$
approaches and extracted from lattice calculations (LC) are presented for comparison.}
\label{tab10}
\begin{tabular}{*{8}{c}}\hline
~$n$~~ & ~~Pad\'e~~ & ~Pad\'e-Borel~ & ~~$R_8$ via $g_8$~~~ & ~3D RG$\cite{SOUK99}$~ &
~$\epsilon$-exp.$\cite{PV2000}$~ & ~LC~ & ~$(1/n)$-exp.~ \\ \hline
~0~~ & ~0.786~ & ~1.614~ & ~$-$0.160~ & ~~      & ~1.1(2)~ & ~~ & ~~ \\
~1~~ & ~0.466~ & ~1.100~ & ~$-$0.008~ & ~0.856~ & ~0.94(14)~ & ~0.871(14)$\cite{BP2011}$~ & ~~ \\
~~  &      &      &      & ~0.857(86)$\cite{GZ97}$~ & ~0.78(5)$\cite{GZ97}$~ &
~0.79(4)$\cite{CPRV2002}$~ \\
~2~~ & ~0.224~ & ~0.726~ & ~0.076~ & ~0.563~ & ~0.71(16) & ~0.494(34)$\cite{CHPRV2001}$~ & ~~ \\
~3~~ & ~0.043~ & ~0.450~ & ~0.124~ & ~ 0.334~ & ~0.33(10) & ~0.21(7)$\cite{CHPRV2002}$~ & ~~ \\
~4~~ & ~$-$0.094~ & ~0.244~ & ~0.134~ & ~0.15~ & ~0.065(80) & ~0.07(14)$\cite{TPV2003}$~ & ~~ \\
~5~~ & ~$-$0.196~ & ~0.089~ & ~0.113~ & ~$-$0.3(9)$\cite{BT2005}$~ & ~$-$0.1(2)$\cite{BT2005}$~ & & \\
~6~~ & ~$-$0.273~ & ~$-$0.029~ & ~0.074~ & ~$-$0.09~ & ~$-$0.2(1)$\cite{BT2005}$~ & &\\
~8~~ & ~$-$0.374~ & ~$-$0.189~ & ~$-$0.022~ & ~$-$0.25~ & ~$-$0.405(31) & ~~ & ~$-$2.885~ \\
~16~~ & ~$-$0.475~ & ~$-$0.391~ & ~$-$0.254~ & ~$-$0.44~ & ~$-$0.528(14) &    & ~$-$1.442~ \\
~32~~ & ~$-$0.398~ & ~$-$0.365~ & ~$-$0.289~ & ~$-$0.42~  & ~$-$0.425(7)~ &  & ~$-$0.721~ \\
~~ &  &  &  & ~$-$0.45(7)$\cite{BT2005}$~ & ~$-$0.427(3)$\cite{BT2005}$~ & \\
~48~~ & ~$-$0.319~ & ~$-$0.301~ & ~$-$0.247~ & ~~ & ~$-$0.322(2)  &     & ~$-$0.481~ \\
~64~~ & ~$-$0.263~ & ~$-$0.252~ & ~$-$0.209~ & ~$-$0.29(3)$\cite{BT2005}$~ &
~$-$0.269(3)$\cite{BT2005}$~ &     & ~$-$0.361~ \\ \hline
\end{tabular}
\end{table}

\begin{table}[t]
\caption{The values of quartic coupling $g^*$ in two dimensions for various $n$
found by Pad\'e and Pad\'e-Borel resummation of corresponding
pseudo-$\epsilon$ expansions and by direct summation of these series with optimal
truncation (OTDS). Pad\'e estimates are those given by approximant [3/2], i. e.
by diagonal approximant [2/2] for $g^*/\tau$. When this approximant has pole close
to 1 approximant [2/3] (marked by subscript) is used. Pad\'e-Borel estimates are
based on approximants free of dangerous -- positive axis -- poles; relevant
approximants are shown as subscripts. For $n = 0$ and $n = 1$ the numbers yielded
by two different working approximants are presented to give an idea about the
level of accuracy of the iteration scheme employed. The fixed point values of
$g$ resulting from 5-loop RG series in two dimensions$\cite{OS2000}$, obtained
within the $\epsilon$-expansion$\cite{PV2000}$ and $(1/n)$-expansion$\cite{CPRV96}$
approaches and extracted from lattice calculations$\cite{BC96,CHPV2001}$ (LC) are
presented for comparison.}
\label{tab11}
\begin{tabular}{*{8}{c}}\hline
~~$n$~~~ & ~~Pad\'e~~ & ~~Pad\'e-Borel~~ & ~~OTDS~~ & ~Constr. $\epsilon$-exp.$\cite{PV2000}$~
& ~~2D RG$\cite{OS2000}$~~ & ~~LC$\cite{BC96}$~~ & ~$(1/n)$-exp.$\cite{CPRV96}$~ \\ \hline
0~~  & ~1.872~ & ~1.862$_{[4/1]}$~ & ~1.831~ & ~1.72(4)~  & ~1.86(4)~  & ~1.676(3)  &  \\
~~   & ~1.749$_{[2/3]}$~ & ~1.710$_{[2/3]}$~ &         &            &            & ~          &  \\
1~~  & ~1.850~ & ~1.839$_{[4/1]}$~ & ~1.897~ & ~1.76(5)~  & ~1.84(3)~  & ~1.7538(5) &  \\
~~   & ~1.751$_{[2/3]}$~ & ~1.710$_{[2/3]}$~ &     &     &     & ~1.754365(3)$\cite{CHPV2001}$~ &  \\
2~~  & ~1.809~ & ~1.799$_{[4/1]}$~ & ~1.845~ & ~1.82(3)~  & ~1.80(3)~  & ~1.82(1)   &  \\
3~~  & ~1.759~ & ~1.751$_{[4/1]}$~ & ~1.787~ & ~1.75(3)~  & ~1.75(2)~  &            & ~1.759~\\
4~~  & ~1.707~ & ~1.712$_{[3/2]}$~ & ~1.729~ & ~1.67(4)~  & ~1.70(2)~  & ~1.66(1)   & ~1.699~\\
8~~  & ~1.531~ & ~1.532$_{[3/2]}$~ & ~1.535~ & ~1.46(3)~  & ~1.52(1)~  & ~1.43(3)   & ~1.480~\\
16~~ & ~1.303$_{[2/3]}$~ & ~1.308$_{[2/3]}$~ & ~1.321~ & ~1.28(2)~  & ~1.313(3)~ &            & ~1.283~\\
24~~ & ~1.213$_{[2/3]}$~ & ~1.219$_{[2/3]}$~ & ~1.206~ & ~1.20(2)~  & ~~         &            & ~1.200~\\
32~~ & ~1.163$_{[2/3]}$~ & ~1.168$_{[2/3]}$~ & ~1.158~ & ~1.16(1)~  & ~1.170(2)~ &            & ~1.154~\\
48~~ & ~1.105~ & ~1.114$_{[2/3]}$~ & ~1.107~ & ~1.11(1)~  & ~~         &            & ~1.106~\\
64~~ & ~1.0806~ & ~1.0859$_{[2/3]}$~ & ~1.0815~ &  ~~        &            &            & ~1.0806~\\
\hline
\end{tabular}
\end{table}

\end{document}